\pdfoutput=1
\documentclass[12pt,a4paper]{article}

\def\figr{0.55}
\def\figrhalf{0.325}

\usepackage{ifthen} 
\newboolean{pdflatex}
\setboolean{pdflatex}{true} 

\newboolean{articletitles}
\setboolean{articletitles}{true} 

\newboolean{uprightparticles}
\setboolean{uprightparticles}{false} 

\usepackage{color}
\definecolor{jblue}  {RGB}{20,50,100}
\definecolor{npurple}  {RGB} {153, 51, 204}
\definecolor{wred}   {RGB}{217,0,56}
\definecolor{white}   {RGB}{255,255,255}

\definecolor{korange}   {RGB}{235, 80,  43}
\definecolor{korange2}   {RGB}{245, 100,  63}
\definecolor{kyelloworange}   {RGB}{255, 210,  110}
\definecolor{kyelloworange2}   {RGB}{240, 170,  90}
\definecolor{kred}   {RGB}{204,  102, 153}
\definecolor{kpurple}   {RGB}{153,  61, 190}
\definecolor{kpurplelight}   {RGB}{213,  161, 230}

\definecolor{sred}   {RGB}{255,  0,  0}
\definecolor{sblue}   {RGB}{0, 0, 255}

\def\paperauthors{LHCb collaboration} 
\def\paperasciititle{Observation of a new excited Ds+ meson in B0->D+D-K+pi- decays} 
\def\papertitle{Observation of a new excited \Ds meson in \BzDDKpi decays} 
\def\paperkeywords{{High Energy Physics}, {LHCb}} 
\def\papercopyright{\the\year\ CERN for the benefit of the LHCb collaboration} 
\def\paperlicence{CC BY 4.0 licence}
\def\paperlicenceurl{https://creativecommons.org/licenses/by/4.0/}

\usepackage[top=1in, bottom=1.25in, left=1in, right=1in]{geometry}

\usepackage{microtype}
\usepackage{lineno}  
\usepackage{xspace} 
\usepackage{caption} 

\usepackage{graphicx}  
\usepackage{color}
\usepackage{colortbl}
\graphicspath{{./figs/}, {./eps/}} 

\usepackage{amsmath} 
\usepackage{amssymb}
\usepackage{amsfonts}
\usepackage{upgreek} 

\newcommand*\patchAmsMathEnvironmentForLineno[1]{%
\expandafter\let\csname old#1\expandafter\endcsname\csname #1\endcsname
\expandafter\let\csname oldend#1\expandafter\endcsname\csname
end#1\endcsname
 \renewenvironment{#1}%
   {\linenomath\csname old#1\endcsname}%
   {\csname oldend#1\endcsname\endlinenomath}%
}
\newcommand*\patchBothAmsMathEnvironmentsForLineno[1]{%
  \patchAmsMathEnvironmentForLineno{#1}%
  \patchAmsMathEnvironmentForLineno{#1*}%
}
\AtBeginDocument{%
\patchBothAmsMathEnvironmentsForLineno{equation}%
\patchBothAmsMathEnvironmentsForLineno{align}%
\patchBothAmsMathEnvironmentsForLineno{flalign}%
\patchBothAmsMathEnvironmentsForLineno{alignat}%
\patchBothAmsMathEnvironmentsForLineno{gather}%
\patchBothAmsMathEnvironmentsForLineno{multline}%
\patchBothAmsMathEnvironmentsForLineno{eqnarray}%
}

\usepackage{hyperxmp}

\usepackage[pdftex,
            pdfauthor={\paperauthors},
            pdftitle={\paperasciititle},
            pdfkeywords={\paperkeywords},
            pdfcopyright={Copyright (C) \papercopyright},
            pdflicenseurl={\paperlicenceurl}]{hyperref}

\usepackage[colorinlistoftodos,textsize=scriptsize]{todonotes}

\usepackage[bottom,flushmargin,hang,multiple]{footmisc}

\usepackage[all]{hypcap} 

\usepackage{xspace} 
\usepackage{upgreek}

\def\lhcb   {\mbox{LHCb}\xspace}

\def\MagUp {\mbox{\em Mag\kern -0.05em Up}\xspace}

\ifthenelse{\boolean{uprightparticles}}%
{

 \def\Pmu         {\ensuremath{\upmu}\xspace}

 \def\Ppi         {\ensuremath{\uppi}\xspace}

 \def\Ppsi        {\ensuremath{\uppsi}\xspace}

 \def\PDelta      {\ensuremath{\Delta}\xspace}                 
 \def\PXi         {\ensuremath{\Xi}\xspace}                 
 \def\PLambda     {\ensuremath{\Lambda}\xspace}                 
 \def\PSigma      {\ensuremath{\Sigma}\xspace}                 
 \def\POmega      {\ensuremath{\Omega}\xspace}                 
 \def\PUpsilon    {\ensuremath{\Upsilon}\xspace}

 \def\PB      {\ensuremath{\mathrm{B}}\xspace}                 
                  
 \def\PD      {\ensuremath{\mathrm{D}}\xspace}

 \def\PJ      {\ensuremath{\mathrm{J}}\xspace}                 
 \def\PK      {\ensuremath{\mathrm{K}}\xspace}

 \def\Pb      {\ensuremath{\mathrm{b}}\xspace}                 
 \def\Pc      {\ensuremath{\mathrm{c}}\xspace}

 \def\Pi      {\ensuremath{\mathrm{i}}\xspace}

 \def\Pp      {\ensuremath{\mathrm{p}}\xspace}

 \def\Ps      {\ensuremath{\mathrm{s}}\xspace}

 \def\thebaroffset{0.0em}
}
{

 \def\Pmu         {\ensuremath{\mu}\xspace}

 \def\Ppi         {\ensuremath{\pi}\xspace}

 \def\Ppsi        {\ensuremath{\psi}\xspace}                 
                  
 \mathchardef\PDelta="7101
 \mathchardef\PXi="7104
 \mathchardef\PLambda="7103
 \mathchardef\PSigma="7106
 \mathchardef\POmega="710A
 \mathchardef\PUpsilon="7107
                  
 \def\PB      {\ensuremath{B}\xspace}                 
                  
 \def\PD      {\ensuremath{D}\xspace}

 \def\PJ      {\ensuremath{J}\xspace}                 
 \def\PK      {\ensuremath{K}\xspace}

 \def\Pb      {\ensuremath{b}\xspace}                 
 \def\Pc      {\ensuremath{c}\xspace}

 \def\Pi      {\ensuremath{i}\xspace}

 \def\Pp      {\ensuremath{p}\xspace}

 \def\Ps      {\ensuremath{s}\xspace}

 \def\thebaroffset{0.18em}
}
\newcommand{\offsetoverline}[2][\thebaroffset]{\kern #1\overline{\kern -#1 #2}}%

\makeatletter
\ifcase \@ptsize \relax
  \newcommand{\miniscule}{\@setfontsize\miniscule{4}{5}}
\or
  \newcommand{\miniscule}{\@setfontsize\miniscule{5}{6}}
\or
  \newcommand{\miniscule}{\@setfontsize\miniscule{5}{6}}
\fi
\makeatother

\DeclareRobustCommand{\optbar}[1]{\shortstack{{\miniscule (\rule[.5ex]{1.25em}{.18mm})}
  \\ [-.7ex] $#1$}}

\def\mumu       {{\ensuremath{\Pmu^+\Pmu^-}}\xspace}

\def\squark    {{\ensuremath{\Ps}}\xspace}

\def\cquark    {{\ensuremath{\Pc}}\xspace}
\def\cquarkbar {{\ensuremath{\overline \cquark}}\xspace}
\def\ccbar     {{\ensuremath{\cquark\cquarkbar}}\xspace}
\def\bquark    {{\ensuremath{\Pb}}\xspace}

\def\pion   {{\ensuremath{\Ppi}}\xspace}
\def\piz    {{\ensuremath{\pion^0}}\xspace}

\def\pim    {{\ensuremath{\pion^-}}\xspace}
\def\pipm   {{\ensuremath{\pion^\pm}}\xspace}

\def\kaon    {{\ensuremath{\PK}}\xspace}

\def\KorKbar {\kern \thebaroffset\optbar{\kern -\thebaroffset \PK}{}\xspace}

\def\Kp      {{\ensuremath{\kaon^+}}\xspace}
\def\Km      {{\ensuremath{\kaon^-}}\xspace}

\def\Kmp     {{\ensuremath{\kaon^\mp}}\xspace}

\def\D       {{\ensuremath{\PD}}\xspace}

\def\DorDbar {\kern \thebaroffset\optbar{\kern -\thebaroffset \PD}\xspace}

\def\Dp      {{\ensuremath{\D^+}}\xspace}
\def\Dm      {{\ensuremath{\D^-}}\xspace}
\def\Dpm     {{\ensuremath{\D^\pm}}\xspace}

\def\DpDm    {\ensuremath{\Dp {\kern -0.16em \Dm}}\xspace}
\def\Dstar   {{\ensuremath{\D^*}}\xspace}

\def\Dstarp  {{\ensuremath{\D^{*+}}}\xspace}

\def\Ds      {{\ensuremath{\D^+_\squark}}\xspace}
\def\Dsp     {{\ensuremath{\D^+_\squark}}\xspace}

\def\B       {{\ensuremath{\PB}}\xspace}

\def\BorBbar {\kern \thebaroffset\optbar{\kern -\thebaroffset \PB}\xspace}
\def\Bz      {{\ensuremath{\B^0}}\xspace}

\def\Bd      {{\ensuremath{\B^0}}\xspace}

\def\BdorBdbar {\kern \thebaroffset\optbar{\kern -\thebaroffset \Bd}\xspace}
\def\Bu      {{\ensuremath{\B^+}}\xspace}

\def\Bs      {{\ensuremath{\B^0_\squark}}\xspace}

\def\BsorBsbar {\kern \thebaroffset\optbar{\kern -\thebaroffset \Bs}\xspace}

\def\jpsi     {{\ensuremath{{\PJ\mskip -3mu/\mskip -2mu\Ppsi}}}\xspace}

\def\Y#1S{\ensuremath{\PUpsilon{(#1S)}}\xspace}

\def\proton      {{\ensuremath{\Pp}}\xspace}

\def\LorLbar     {\kern \thebaroffset\optbar{\kern -\thebaroffset \PLambda}\xspace}

\newcommand{\decay}[2]{\ensuremath{#1\!\to #2}\xspace} 

\def\to                 {\ensuremath{\rightarrow}\xspace}

\def\AT#1     {\ensuremath{A_{\mathrm{T}}^{#1}}\xspace}           

\def\C#1      {\ensuremath{\mathcal{C}_{#1}}\xspace}                       
\def\Cp#1     {\ensuremath{\mathcal{C}_{#1}^{'}}\xspace}                    
\def\Ceff#1   {\ensuremath{\mathcal{C}_{#1}^{\mathrm{(eff)}}}\xspace}        
\def\Cpeff#1  {\ensuremath{\mathcal{C}_{#1}^{'\mathrm{(eff)}}}\xspace}       
\def\Ope#1    {\ensuremath{\mathcal{O}_{#1}}\xspace}                       
\def\Opep#1   {\ensuremath{\mathcal{O}_{#1}^{'}}\xspace}                    

\newcommand{\aunit}[1]{\ensuremath{\text{\,#1}}}       

\newcommand{\tev}{\aunit{Te\kern -0.1em V}\xspace}
\newcommand{\gev}{\aunit{Ge\kern -0.1em V}\xspace}
\newcommand{\mev}{\aunit{Me\kern -0.1em V}\xspace}
\newcommand{\kev}{\aunit{ke\kern -0.1em V}\xspace}
\newcommand{\ev}{\aunit{e\kern -0.1em V}\xspace}
\newcommand{\gevgev}{\ensuremath{\gev^2}\xspace} 
\newcommand{\mevc}{\ensuremath{\aunit{Me\kern -0.1em V\!/}c}\xspace}
\newcommand{\gevc}{\ensuremath{\aunit{Ge\kern -0.1em V\!/}c}\xspace}
\newcommand{\mevcc}{\ensuremath{\aunit{Me\kern -0.1em V\!/}c^2}\xspace}
\newcommand{\gevcc}{\ensuremath{\aunit{Ge\kern -0.1em V\!/}c^2}\xspace}

\def\fm   {\aunit{fm}\xspace}

\def\fb   {\ensuremath{\aunit{fb}}\xspace}
\def\invfb   {\ensuremath{\fb^{-1}}\xspace}

\newcommand{\stat}{\aunit{(stat)}\xspace}
\newcommand{\syst}{\aunit{(syst)}\xspace}

\newcommand{\chisq}{\ensuremath{\chi^2}\xspace}

\def\gsim{{~\raise.15em\hbox{$>$}\kern-.85em
          \lower.35em\hbox{$\sim$}~}\xspace}
\def\lsim{{~\raise.15em\hbox{$<$}\kern-.85em
          \lower.35em\hbox{$\sim$}~}\xspace}

\def\sPlot{\mbox{\em sPlot}\xspace}

\def\mrad{\aunit{mrad}\xspace}

\def\tell1  {TELL1\xspace}
\def\ukl1   {UKL1\xspace}

\newcommand{\phz}{\phantom{0}}

\usepackage{cite} 
\usepackage{mciteplus}

\def\pp{\ensuremath{pp}\xspace}
\def\swave{$S$-wave\xspace}

\def\Mp{\ensuremath{m_R}\xspace}
\def\Gp{\ensuremath{\Gamma_R}\xspace}

\def\DsS{\ensuremath{\D_s(2^1S_0)^+}\xspace}

\def\Dsk{\ensuremath{\D_{sk}}\xspace}
\def\DsJ{\ensuremath{{\D_{sJ}^+}}\xspace}
\def\Dszero{\ensuremath{{\D_{s0}^+}}\xspace}
\def\Dsone{\ensuremath{{\D_{s1}^+}}\xspace}
\def\Dso{\mbox{\ensuremath{\D_{s1}(2536)^+}}\xspace}
\def\Dsn{\mbox{\ensuremath{\D_{s0}(2590)^+}}\xspace}
\def\Kap{\mbox{\ensuremath{K^*_0(700)^0}}\xspace}

\def\DKpi{\mbox{\ensuremath{\Dp\Kp\pim}}\xspace}
\def\Kpi{\mbox{\ensuremath{\Kp\pim}}\xspace}

\def\Pars{\ensuremath{\overrightarrow{\omega}}}
\def\mDKpi{\ensuremath{{m_{\Dp\Kp\pim}}\xspace}}
\def\mKpi{\ensuremath{{m_{\Kp\pim}}\xspace}}

\def\Like {{\cal L}}

\newcommand{\BzDDKpi}{\mbox{\ensuremath{\Bz\to\Dm\Dp\Kp\pim}}\xspace}

\newcommand{\phdz}{\phantom{.0}}

\usepackage{longtable} 

\begin{document}

\renewcommand{\thefootnote}{\fnsymbol{footnote}}
\setcounter{footnote}{1}


\begin{titlepage}
\pagenumbering{roman}

\vspace*{-1.5cm}
\centerline{\large EUROPEAN ORGANIZATION FOR NUCLEAR RESEARCH (CERN)}
\vspace*{1.5cm} \noindent
\begin{tabular*}{\linewidth}{lc@{\extracolsep{\fill}}r@{\extracolsep{0pt}}}
\ifthenelse{\boolean{pdflatex}}
{\vspace*{-1.5cm}\mbox{\!\!\!\includegraphics[width=.14\textwidth]{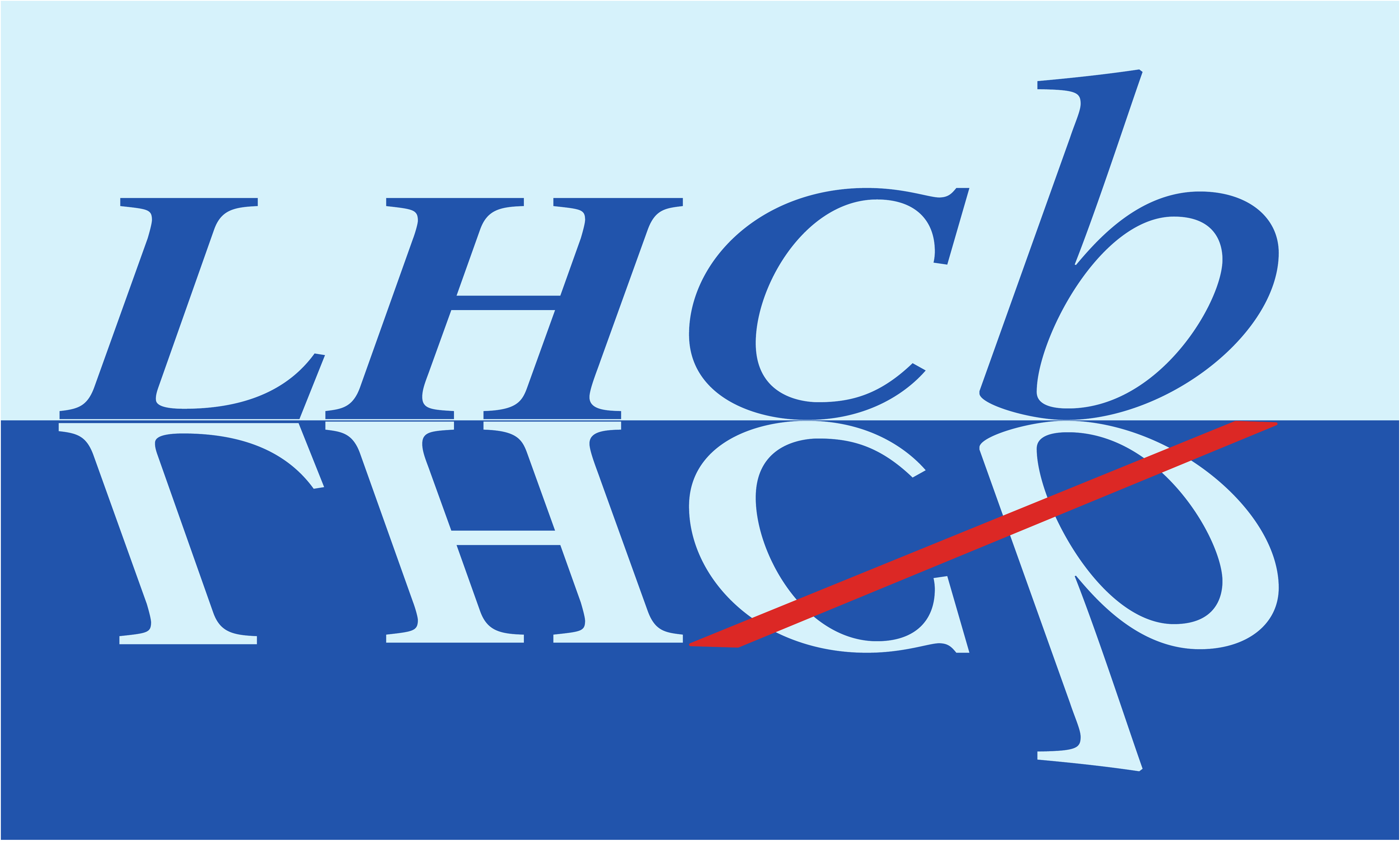}} & &}%
{\vspace*{-1.2cm}\mbox{\!\!\!\includegraphics[width=.12\textwidth]{figs/lhcb-logo.eps}} & &}%
\\
 & & CERN-EP-2020-203 \\  
 & & LHCb-PAPER-2020-034 \\  
 & & March 26, 2021\\ 
 & &  \\
\end{tabular*}

\vspace*{4.0cm}

{\normalfont\bfseries\boldmath\huge
\begin{center}
  \papertitle 
\end{center}
}

\vspace*{2.0cm}

\begin{center}
\paperauthors\footnote{Authors are listed at the end of this Letter.}
\end{center}

\vspace{\fill}

\begin{abstract}
  \noindent
  Using \pp collision data corresponding to an integrated luminosity of 5.4\invfb collected with the \lhcb detector at a center-of-mass energy of 13\tev, the \BzDDKpi decay is studied. A new excited \Ds meson is observed decaying into the \DKpi final state with large statistical significance.
  The pole mass and width, and the spin parity of the new state are measured with an amplitude analysis to be 
  $\Mp=2591\pm6\pm7\mev$, 
  $\Gp=89\pm16\pm12\mev$ and $J^P=0^-$, where the first uncertainty is statistical and the second systematic. 
  Fit fractions for all components in the amplitude analysis are also reported. 
  The new resonance, denoted as \Dsn, is a strong candidate to be the \DsS state, the radial excitation of the pseudoscalar ground-state \Ds meson.
\end{abstract}

\vspace*{2.0cm}

\begin{center}
  Published in Phys. Rev. Lett. 126 (2021) 122002
\end{center}

\vspace{\fill}

{\footnotesize 
\centerline{\copyright~\papercopyright. \href{\paperlicenceurl}{\paperlicence}.}}
\vspace*{2mm}

\end{titlepage}


\newpage
\setcounter{page}{2}
\mbox{~}
%
%
%
%


\renewcommand{\thefootnote}{\arabic{footnote}}
\setcounter{footnote}{0}

\cleardoublepage


\pagestyle{plain} 
\setcounter{page}{1}
\pagenumbering{arabic}


Charm meson spectroscopy is of great theoretical and experimental interest as a testing ground for models based on quantum chromodynamics~(QCD).
In particular, the spectrum of charm-strange mesons has drawn particular attention since the discoveries of the \mbox{$D_{s0}^*(2317)^+$} and \mbox{$D_{s1}(2460)^+$} resonances~\cite{Aubert:2003fg,Besson:2003cp}, with masses much smaller than those predicted for $c\bar{s}$ mesons~\cite{Godfrey:2015dva}.
Interpretations of these states as compact $[cq][\bar{s}\bar{q}]$ tetraquarks~\cite{PhysRevD.71.014028, BROWDER2004365} or $D^{(*)}K$ molecules~\cite{PhysRevD.68.054006,PhysRevLett.91.012003} have been proposed. 
Recent evidence for exotic mesons containing $cs$ rather than $c\bar{s}$ quarks~\cite{LHCb-PAPER-2020-024,LHCb-PAPER-2020-025}, has raised further interest in the interpretation of the \mbox{$D_{s0}^*(2317)^+$} and \mbox{$D_{s1}(2460)^+$} states.  Additional experimental input on the spectrum of $c\bar{s}$ mesons is essential to solve this puzzle.

Meson states are characterized by $n^{2S+1}L_{J}$, and grouped according to $nL$, where $n$ is the principal quantum number, $L$ is the orbital angular momentum between the constituent quarks~($S, P, D$ correspond to $L=0,1,2$), $S=0$ or $1$ is the sum of quark spins and $J$ is the total spin of the meson. 
In the charm-strange meson system, candidates for the two $1S$ mesons and the four $1P$ states are experimentally well-established~\cite{PDG2020}.
Candidates for two of the four $1{D}$ states have also been reported, but their properties need further experimental confirmation~\cite{Palano:2020ifu}.
Only one radial excitation, the $2^3{S}_1$ state $D_{s1}^*(2700)^+$, is currently known.
Among the missing resonances, the $2^1{S}_0$ state, the radial excitation of the pseudoscalar ground-state \Ds\ meson, is expected to be the lightest, with mass around $2.6\gev$. 
Natural units with $\hbar=c=1$ are used, and the inclusion of charge-conjugate processes is implied throughout this Letter.

Studies of \B-meson decays have proven to provide excellent potential to discover new charm-strange mesons and measure their properties~\cite{Brodzicka:2007aa,Lees:2014abp,LHCb-PAPER-2014-035,LHCb-PAPER-2014-036}.
Most such studies to date, however, only address excited \Ds mesons decaying into a $DK$ pair, and hence are only sensitive to \Ds states with natural spin parity ($J^P = 0^+, 1^-, 2^+, ...$) due to parity conservation in strong decays.
The possibility to study production in \B decays of \Ds resonances decaying to the \DKpi final state has not been explored previously, providing opportunities to discover states with masses above $2.5\gev$.
The \Kpi system can be assumed to be in \swave ($J^P = 0^+$) if its mass is restricted to be below the threshold for ${K^{*}(892)^0}$ production.
In this case only \Ds resonances with unnatural spin parity ($J^P = 0^-, 1^+, 2^-, ...$) can decay to \DKpi.

In this Letter, the observation of a new excited \Ds state in the \DKpi mass spectrum is presented.
The results are obtained from an amplitude analysis of \BzDDKpi decays, where the \Kpi mass is restricted to be lower than $0.75\gev$, referred to hereafter as the low \Kpi mass region. 
The analysis makes use of the \pp collision data collected by the \lhcb experiment from 2016 to 2018 at a center-of-mass energy of $13\tev$, corresponding to an integrated luminosity of 5.4\invfb. 

The LHCb detector~\cite{LHCb-DP-2008-001,LHCb-DP-2014-002} is a single-arm forward spectrometer covering the pseudorapidity range $2 < \eta < 5$, designed for the study of particles containing \bquark\ or \cquark quarks.
The on-line event selection is performed by a trigger that consists of a hardware stage, based on information from the calorimeter and muon systems, followed by a software stage, which applies a full event reconstruction~\cite{LHCb-DP-2012-004,LHCb-PROC-2015-018}. 
The momentum scale is calibrated using samples of $\decay{\jpsi}{\mumu}$ and $\decay{\Bu}{\jpsi\Kp}$ decays collected concurrently with the data sample used for this analysis~\cite{LHCb-PAPER-2012-048,LHCb-PAPER-2013-011}. 
Simulated samples are produced with the software packages described in Refs.~\cite{Sjostrand:2007gs,LHCb-PROC-2010-056,Lange:2001uf, Golonka:2005pn,Allison:2006ve,LHCb-PROC-2011-006} and are used to model the effects of the detector acceptance and the imposed selection requirements.

Signal \Bz candidates are formed using \BzDDKpi decays with \Dpm\ candidates reconstructed in the \mbox{$\Kmp\pipm\pipm$} final state.
All final-state particles are required to have particle-identification information consistent with their respective mass hypotheses, 
and to be inconsistent with originating from any primary \proton\proton collision vertex~(PV). The opening angle between any two of the final-state particles is required to be larger than $0.5\mrad$ to suppress potential background from repeated use of track segments.
The \Dpm candidates are required to have good vertex-fit quality and mass within $\pm25\mev$ of the known value~\cite{PDG2020}. 
The decay vertex of the \Bz\ candidate is required to be well reconstructed, to be significantly displaced from all PVs, and to be on a trajectory consistent with having originated from the associated PV.
Both \Dp and \Dm vertices are required to be significantly displaced from the \Bz vertex to suppress contributions from \Bz decays involving one or no \Dpm\ mesons but having the same set of final tracks. 
A kinematic fit is applied to the decay chain to improve the \Bz mass resolution, requiring the \Bz candidate to originate from the associated PV and constraining the masses of the \Dpm candidates to their known values~\cite{PDG2020}. 
The \Bz-candidate mass is additionally constrained to the known value~\cite{PDG2020} in the amplitude analysis. 
For events with multiple \Bz candidates, only that with the lowest kinematic-fit \chisq\ is retained.

Background contributions from misidentification of a final-state pion, kaon or proton in a \bquark-hadron decay to the \mbox{$\Dm\Dp h^+h^{\prime -}$} final state with $h^{(\prime)} \in \left( \pion, \kaon, \proton \right)$ are from Cabibbo-suppressed processes and thus negligible.
An exception is the \mbox{$\Bs\to\Dm\Dp\Kp\Km$} decay, which instead is suppressed by the ratio of fragmentation fractions $f_s/f_d$~\cite{LHCb-PAPER-2018-050,LHCb-PAPER-2019-020} 
and the lack of expected contributions from any charm or charm-strange resonances. 
Partially reconstructed backgrounds with a missing soft neutral pion from the \mbox{$\Dstarp\to\Dp\piz$} decay are also possible but fall below the considered \Bz-candidate mass window of $\pm 100 \mev$ around the known \Bz mass~\cite{PDG2020}. 
Partially reconstructed background involving $\Dstarp\to\Dp\gamma$ decay could have a tail that enters the mass window but is suppressed by its low branching fraction~\cite{PDG2020}.
Hence the only significant source of background that passes the selection is due to random combinations of particles.  

\begin{figure}[b]
    \centering
    \includegraphics[width=\figr\textwidth]{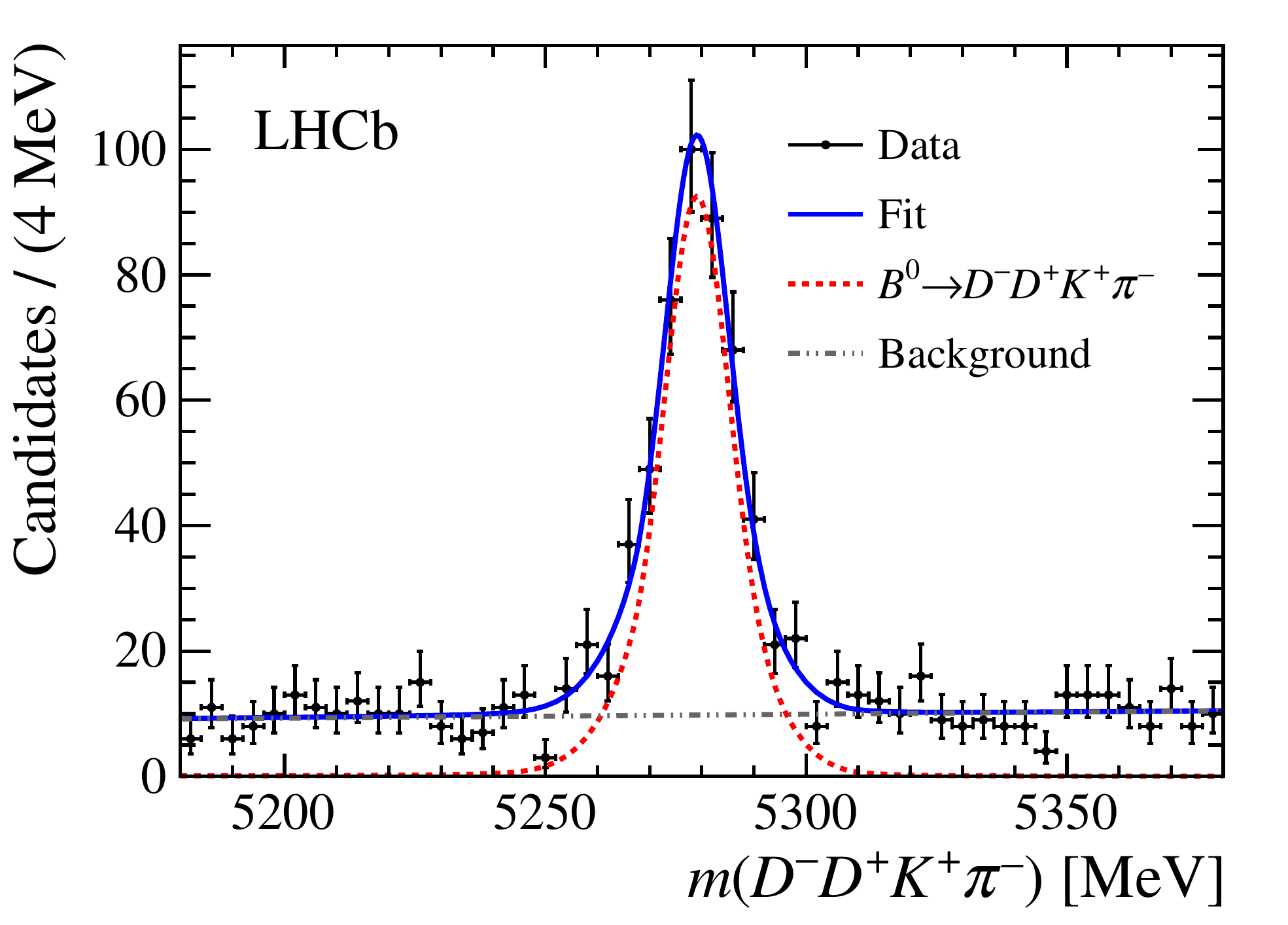}
    \caption{Mass distribution of the selected \Bz candidates in the low \Kp\pim mass region. 
    The fit result is overlaid. 
}
    \label{fig:fitmB}
\end{figure}

\begin{figure}[!btp]
    \centering
    \includegraphics[width=\figr\textwidth]{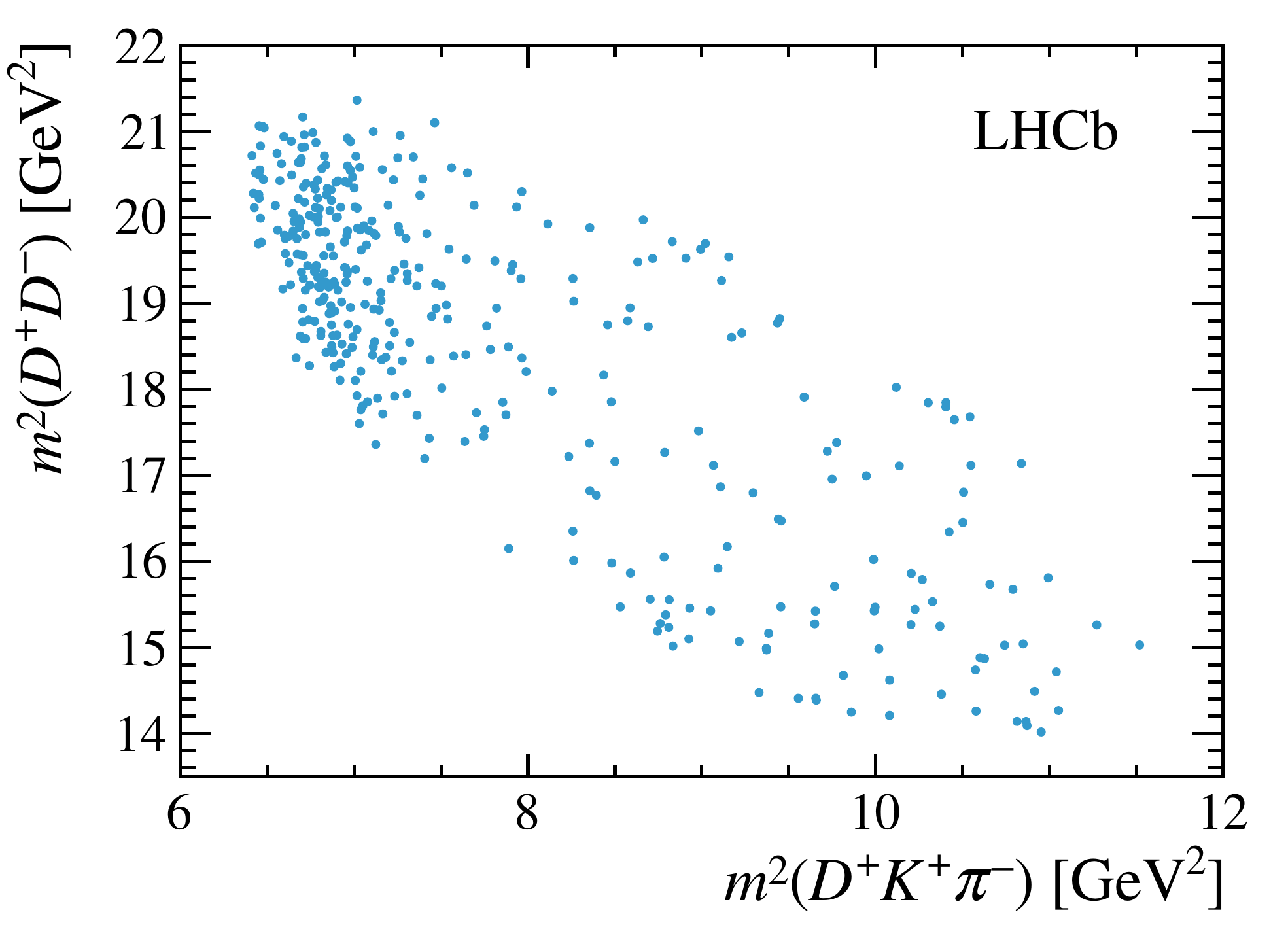}
    \caption{Dalitz plot of the $\Dp\Dm$ versus \DKpi masses-squared for \Bz candidates with  masses within $\pm20\mev$ around the known \Bz~mass~\cite{PDG2020} in the low \Kpi mass region.
}
    \label{fig:dlz}
\end{figure}

\begin{figure}[!bt]
    \centering
    \includegraphics[width=\figrhalf\textwidth]{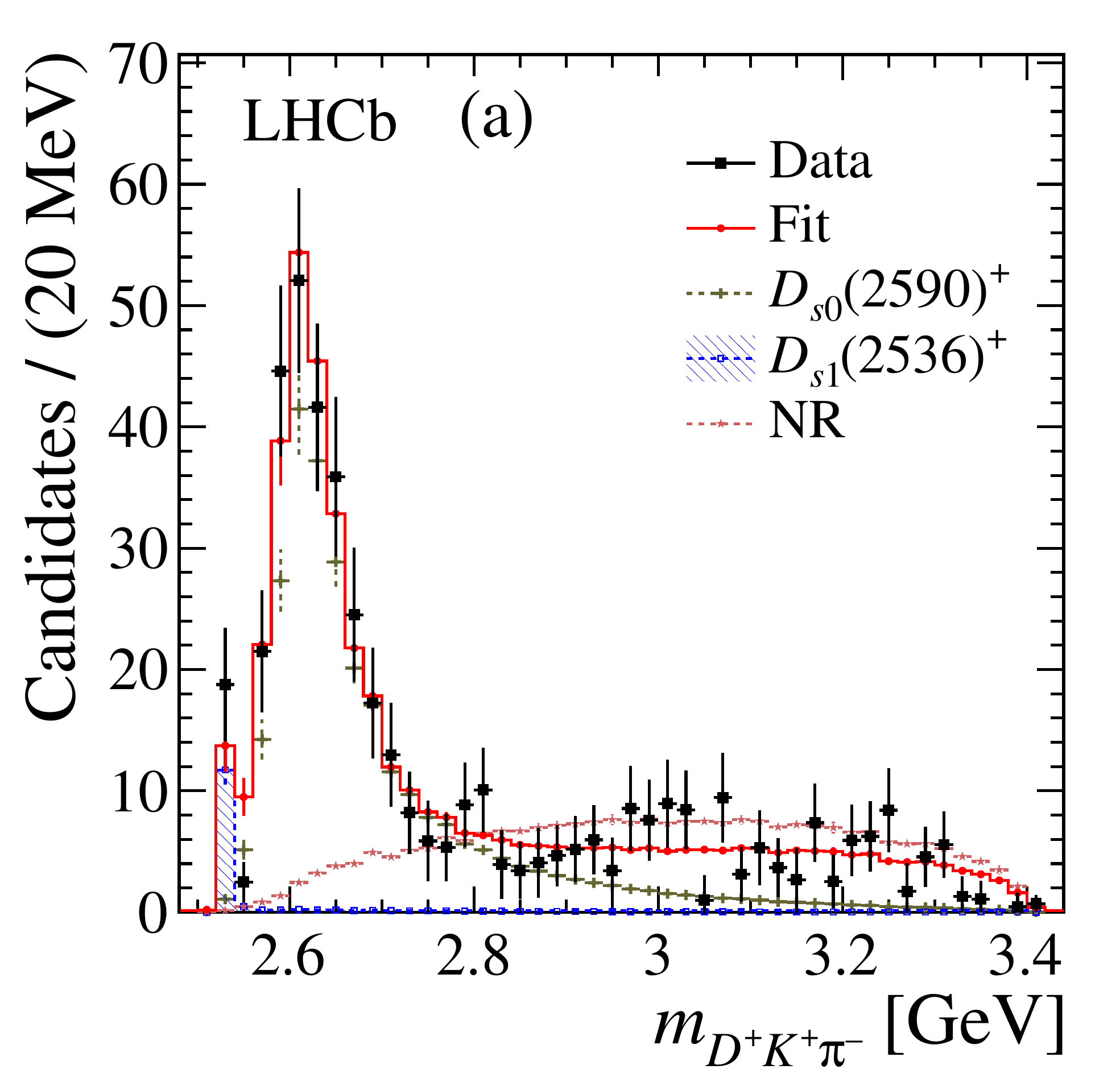}
    \includegraphics[width=\figrhalf\textwidth]{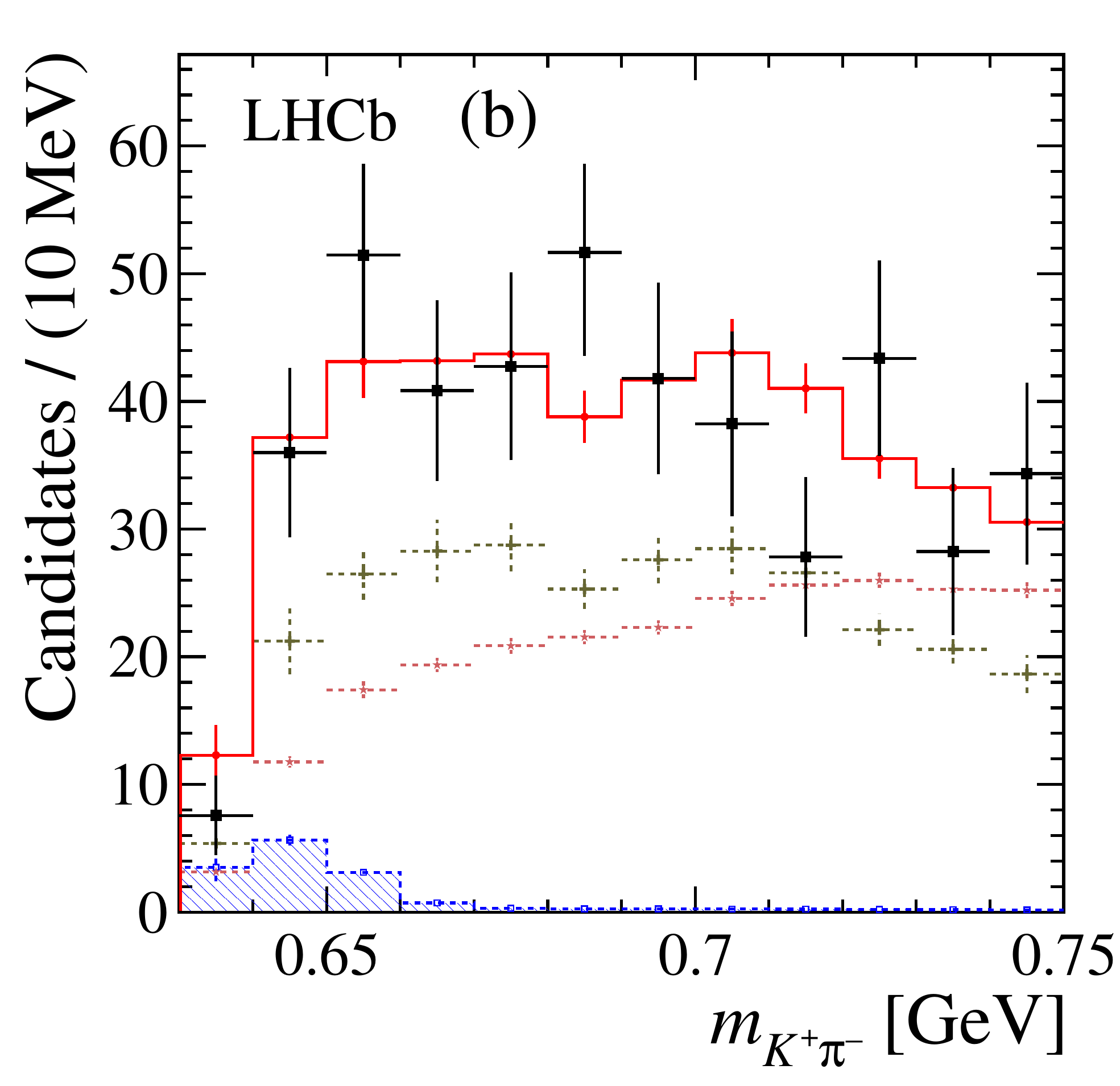}\\
    \includegraphics[width=\figrhalf\textwidth]{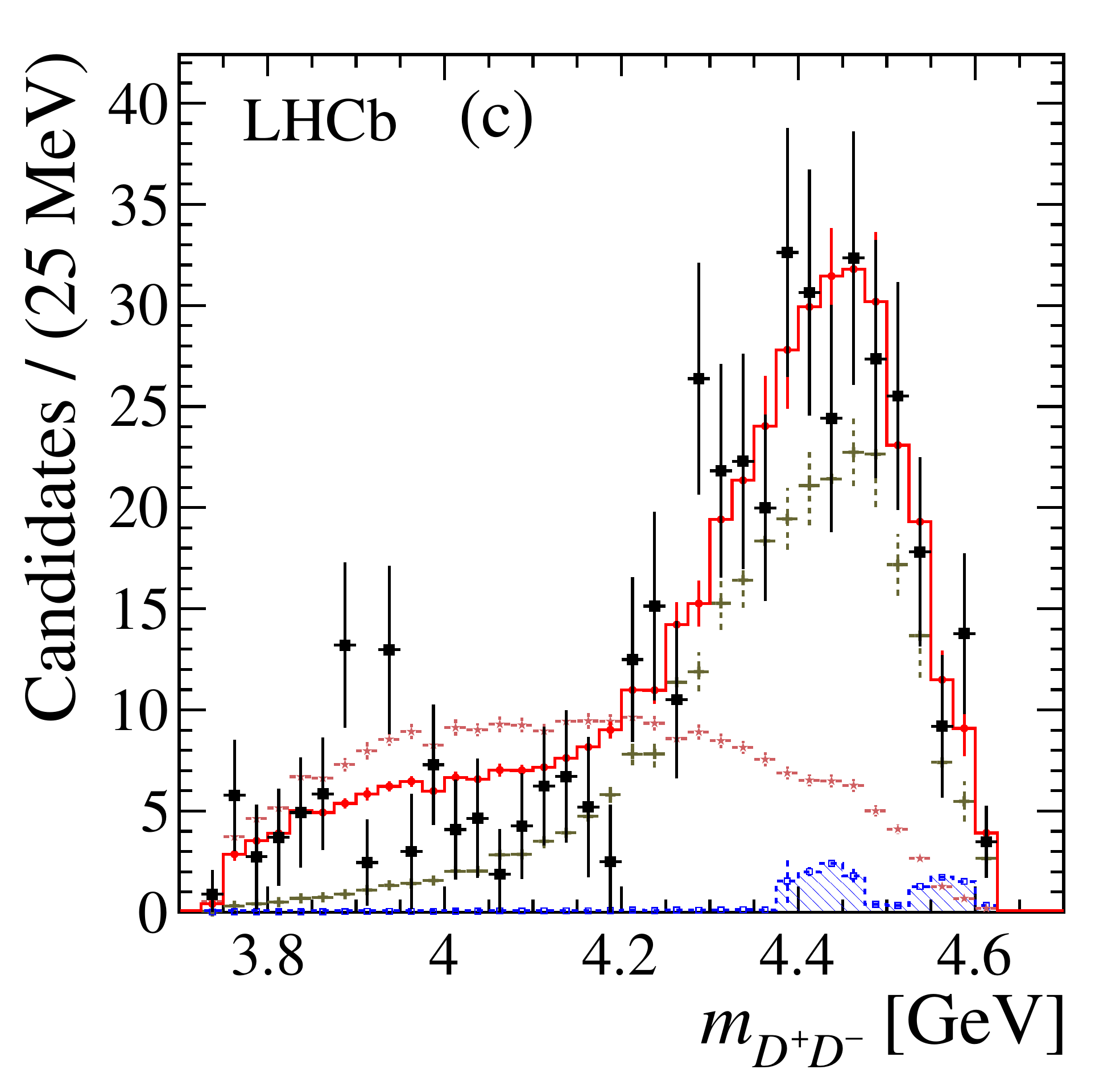}
    \includegraphics[width=\figrhalf\textwidth]{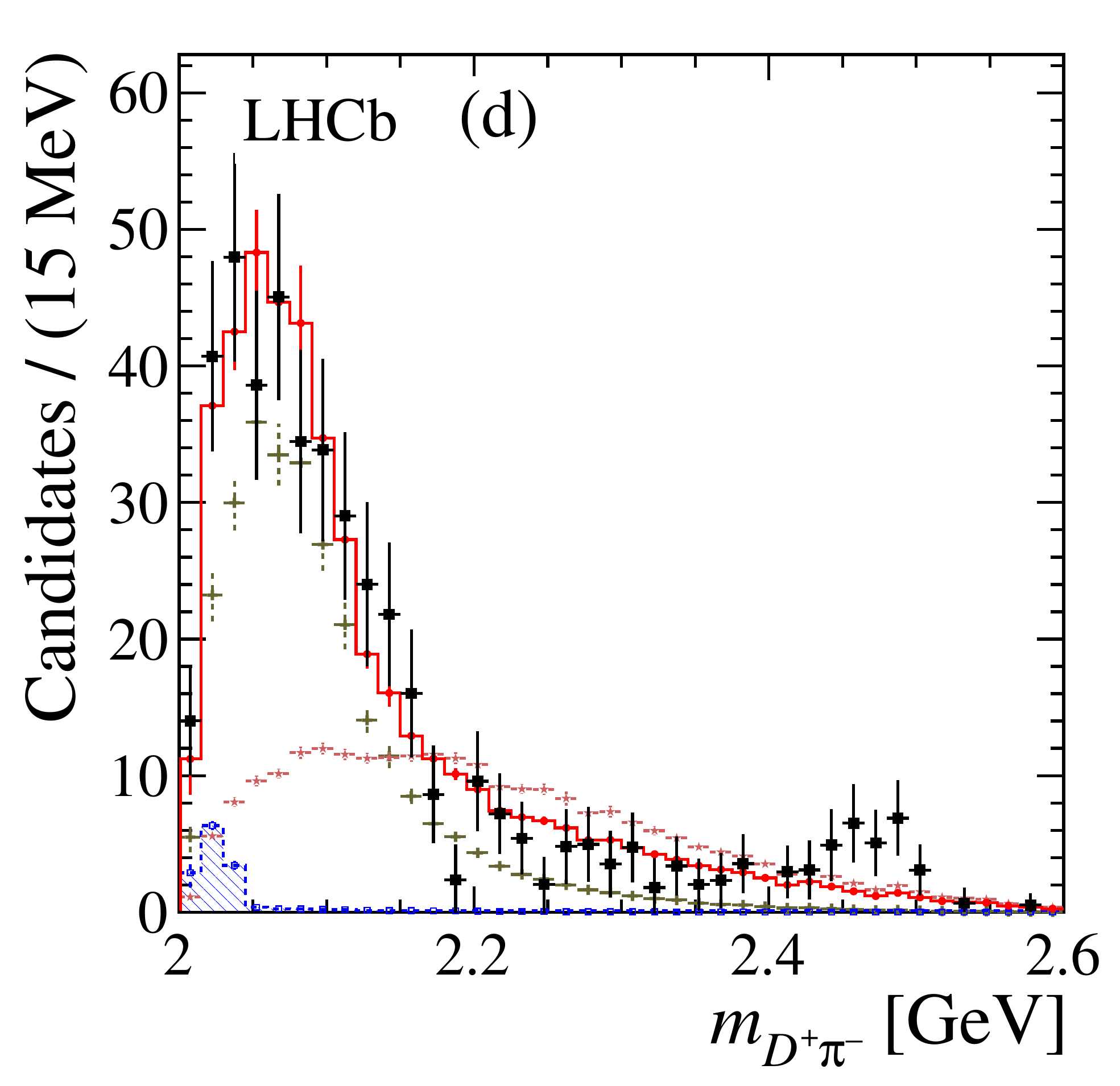}
    \caption{Mass projections of (a)~\DKpi, (b)~\Kpi, (c)~\Dp\Dm and (d)~\Dp\pim systems. 
    Data points are shown in black with the background subtracted statistically using the \sPlot method~\cite{Pivk:2004ty}. 
    Results of the fit with the \Dsn ($J^P=0^-$) model are overlaid as a solid red histogram, and individual contributions shown as dotted histograms.}
    \label{fig:projection}
\end{figure}

An unbinned maximum-likelihood fit is performed to the mass distribution of the \Bz candidates in the low \Kp\pim mass region shown in Fig.~\ref{fig:fitmB}. 
The signal is modeled by a sum of two Crystal Ball functions~\cite{Skwarnicki:1986xj} with a common mean and opposite-side tails. The background is modeled by an exponential function. 
The \Bz signal yield is determined to be $444\pm27$, where the uncertainty is statistical. 
The Dalitz plot~\cite{Dalitz:1953cp,*Fabri:1954zz} of the $\Dp\Dm$ versus \DKpi masses squared for candidates with  masses within $20\mev$ of the known \Bz~mass~\cite{PDG2020} is shown in Fig.~\ref{fig:dlz}. 
A clear cluster of candidates is observed in the \DKpi mass at $\sqrt{6.8\gevgev}\approx2.6\gev$. No $\Dp\Dm$ resonant structure is apparent. 
The \DKpi mass projection is shown in Fig.~\ref{fig:projection}~(a), where a structure at about $2.6\gev$, which has never been observed before, is evident and the small peak just above threshold corresponds to the $D_{s1}(2536)^+$ state~\cite{Balagura:2007dya}.

An amplitude analysis is employed to study structures in the \Dp\Kp\pim system of \Bz\ candidates in the low \Kpi mass region. 
Three \Ds components with unnatural spin parity are considered: a new \Ds state at about $2.6\gev$ denoted hereafter as $\DsJ$ due to its undetermined spin parity, the $J^P=1^+$ \Dso state, and a $J^P=0^-$ nonresonant~(NR) component. 
The line shape of the \Kpi system is modeled by the $J^P=0^+$ \Kap state for all three \Dsp components. 
The amplitude is constructed using the helicity formalism~\cite{Chung:1971ri,*Richman:1984gh,*Jacob:1959at}, with the total amplitude given by 
\begin{equation*}
   {\cal M} =  \sum_k {\cal H}^{\Dsk} \, d^{J_{\Dsk}}_{0,0}(\theta_{D_s}) \,
              p^{L_{\Bz}} \, F_{L_\Bz}(pa)\, q^{L_{\Dsk}} \, F_{L_{\Dsk}}(qa)\, 
                {\rm BW}(\mKpi) \, {\rm BW}_{\Dsk}(\mDKpi),
\end{equation*}
where the summation is over the three \Ds components. 
Here ${\cal H}^{\Dsk}$ is the complex helicity coupling parameter describing the magnitude and the phase of the \Dsk component, and $d^{J_{\Dsk}}_{0,0}(\theta_{D_s})$ is the Wigner small-$d$ matrix with the two subscripts set to zero and the superscript corresponding to the spin of the \Dsk component, where $\theta_{D_s}$ is the angle between the directions of \Dp momentum and the opposite of the \Bz momentum, both in the \Ds rest frame. 
The quantity $p~(q)$ is the momentum of the decay products of the \Bz~(\Dsk) state in its rest frame, and $L_\Bz~(L_{\Dsk})$ is the orbital angular momentum between the decay products of the \Bz~(\Dsk) state. 
The function $F_L(z)$ is the Blatt-Weisskopf form factor that accounts for the barrier of the decay~\cite{Blatt:1952ije,VonHippel:1972fg,Chung:1995dx}, in which $z \equiv pa$ or $qa$ and the parameter $a$ describes the size of the decaying particle, set to \mbox{$3\gev^{-1}\sim0.6\fm$}. 
The line shapes of the \DsJ, \Dso, and \Kap states are described by relativistic Breit-Wigner~(BW) functions.
The NR component has a constant line shape. 

Different expressions for the width $\Gamma(m)$ that enters the BW function are used for the \DsJ, \Dso and \Kap states. 
The \Dso width is set to constant as it is very narrow, while a two-body mass-dependent width is used for the \Kap\ state; in both cases the BW parameters are fixed to their known values~\cite{Lees:2011um,PDG2020}.
The total \DsJ width is described as the sum of contributions from the open decay channels to two-body $\Dstar\kaon$ and three-body $\D\kaon\pi$ decays, 
\begin{equation*}
\Gamma^\DsJ(\mDKpi)=\Gamma^{\DsJ\to\Dstar\kaon}(\mDKpi)+\Gamma^{\DsJ\to\D\kaon\pi}(\mDKpi)\,,
\end{equation*} 
where \mbox{$\Gamma^{\DsJ\to\Dstar\kaon}$} and \mbox{$\Gamma^{\DsJ\to\D\kaon\pi}$} are the partial widths for the corresponding decays.
The former is parameterized with a two-body mass-dependent width and the latter is set to a constant. 

The signal model in the amplitude analysis is the amplitude squared $|{\cal M}|^2$ multiplied by an efficiency function and normalised to unity when integrated over the phase space. 
The unknown parameters of the signal model, denoted hereafter as \Pars, are the complex helicity coupling parameters of the \DsJ and \Dso states, the BW mass and width of the \DsJ state, and the width fraction of the $\DsJ\to\Dstar\kaon$ channel defined as \mbox{$r={\Gamma^{\DsJ\to\D\kaon\pi}(m_0) / \Gamma^\DsJ}(m_0)$}, where $m_0$ is the BW mass.
The helicity coupling parameter of the NR component is fixed to unity to serve as a reference amplitude.
The optimal values of the parameters, $\Pars_{\rm min}$, are determined with the same method used, and described in detail in Ref.~\cite{LHCb-PAPER-2015-029}.
An unbinned fit minimizes the negative log-likelihood, $-2\ln\Like(\Pars)$, with the background subtracted statistically using weights obtained with the \sPlot method with the \Bz-candidate mass as the discriminating variable. 
The variables in the amplitude analysis, $\mDKpi$, $\mKpi$ and $\theta_{D_s}$, are confirmed not to have strong correlations with the \Bz-candidate mass, as required in the \sPlot method. 
The non-parametric efficiency function is determined from simulation with corrections applied to ensure the trigger efficiency, \Bz kinematics and track multiplicity match those observed in data. 

\begin{figure}[!bt]
    \centering
    \includegraphics[width=\figrhalf\textwidth]{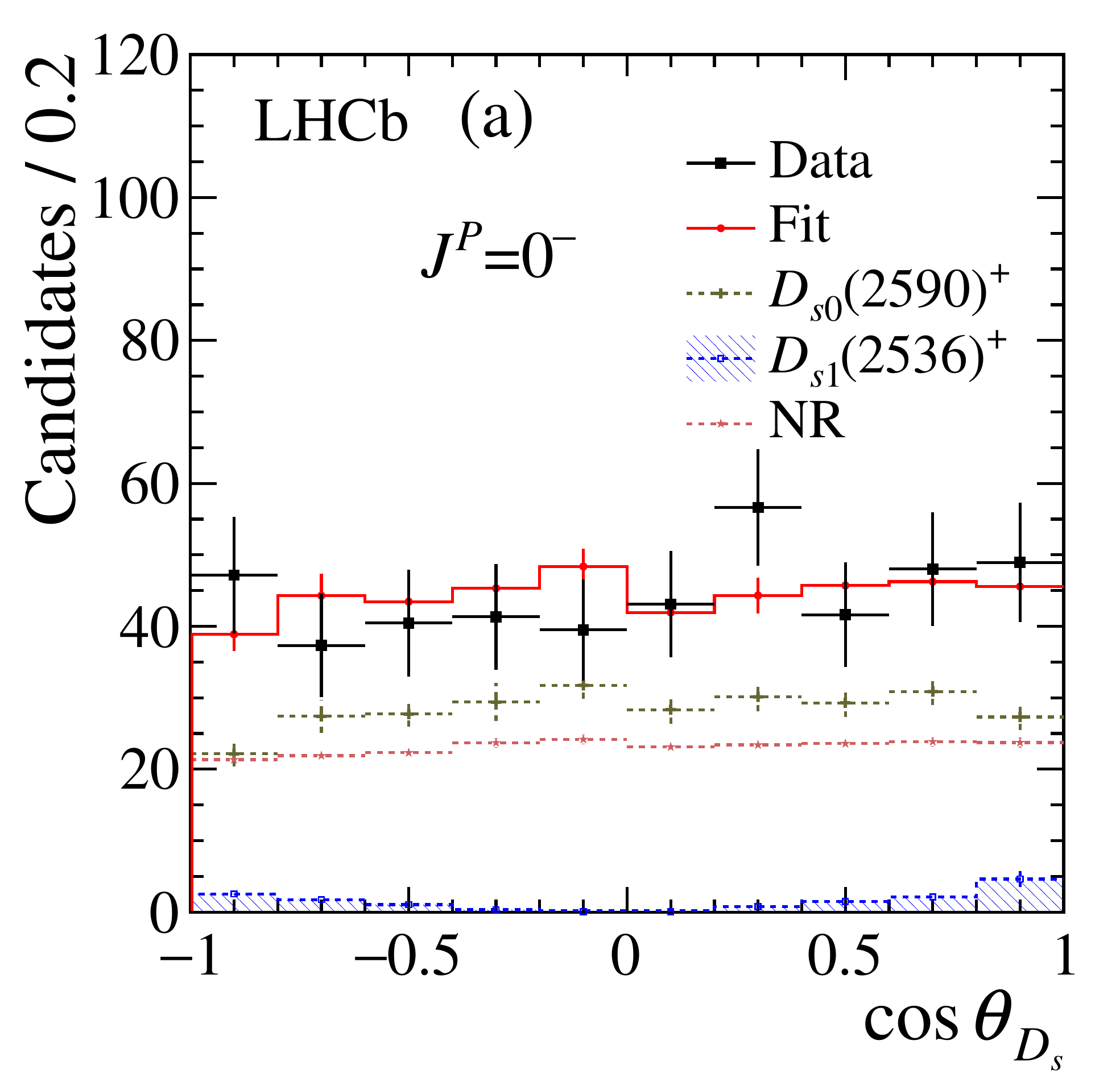}
    \includegraphics[width=\figrhalf\textwidth]{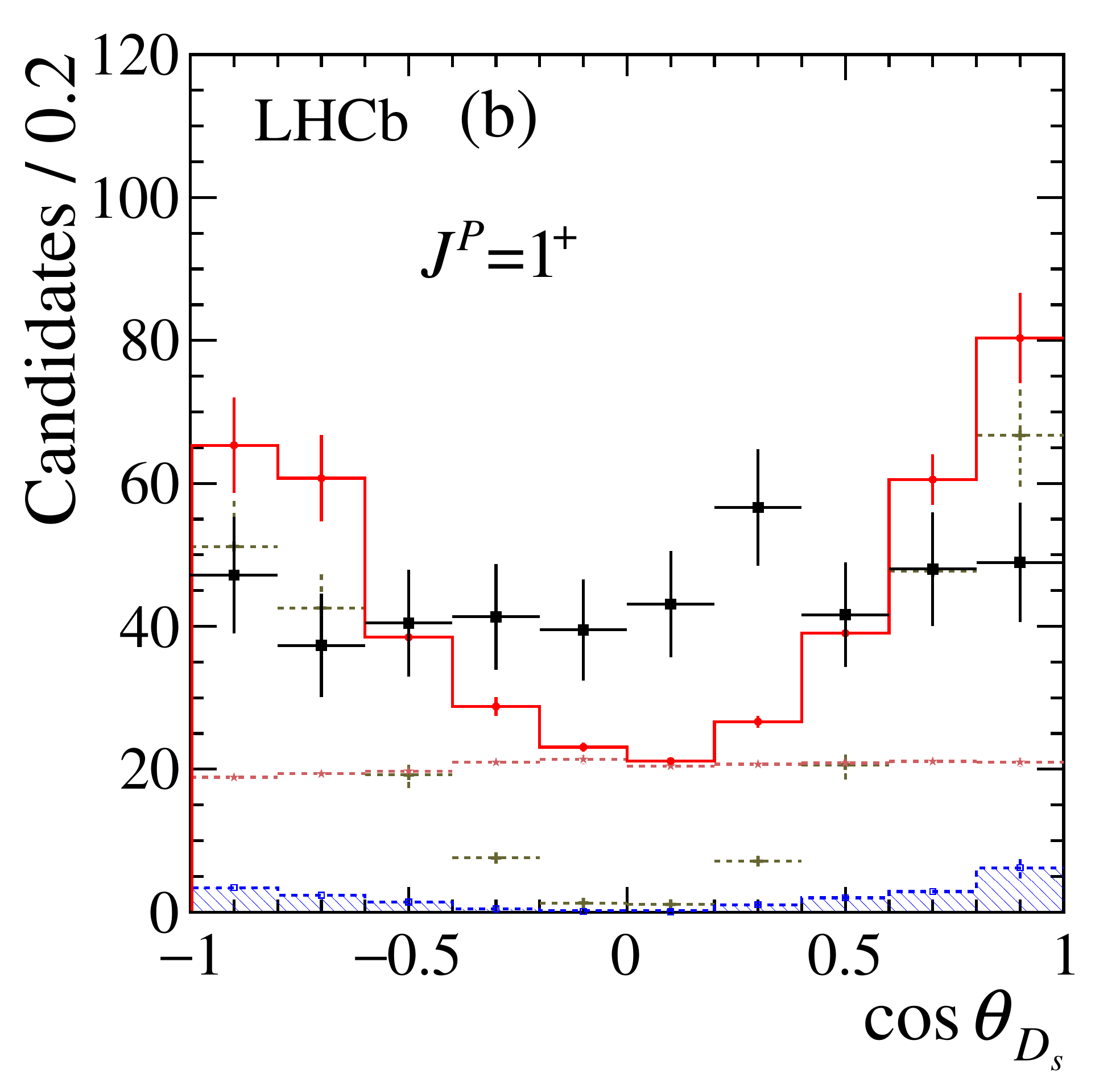}
    \includegraphics[width=\figrhalf\textwidth]{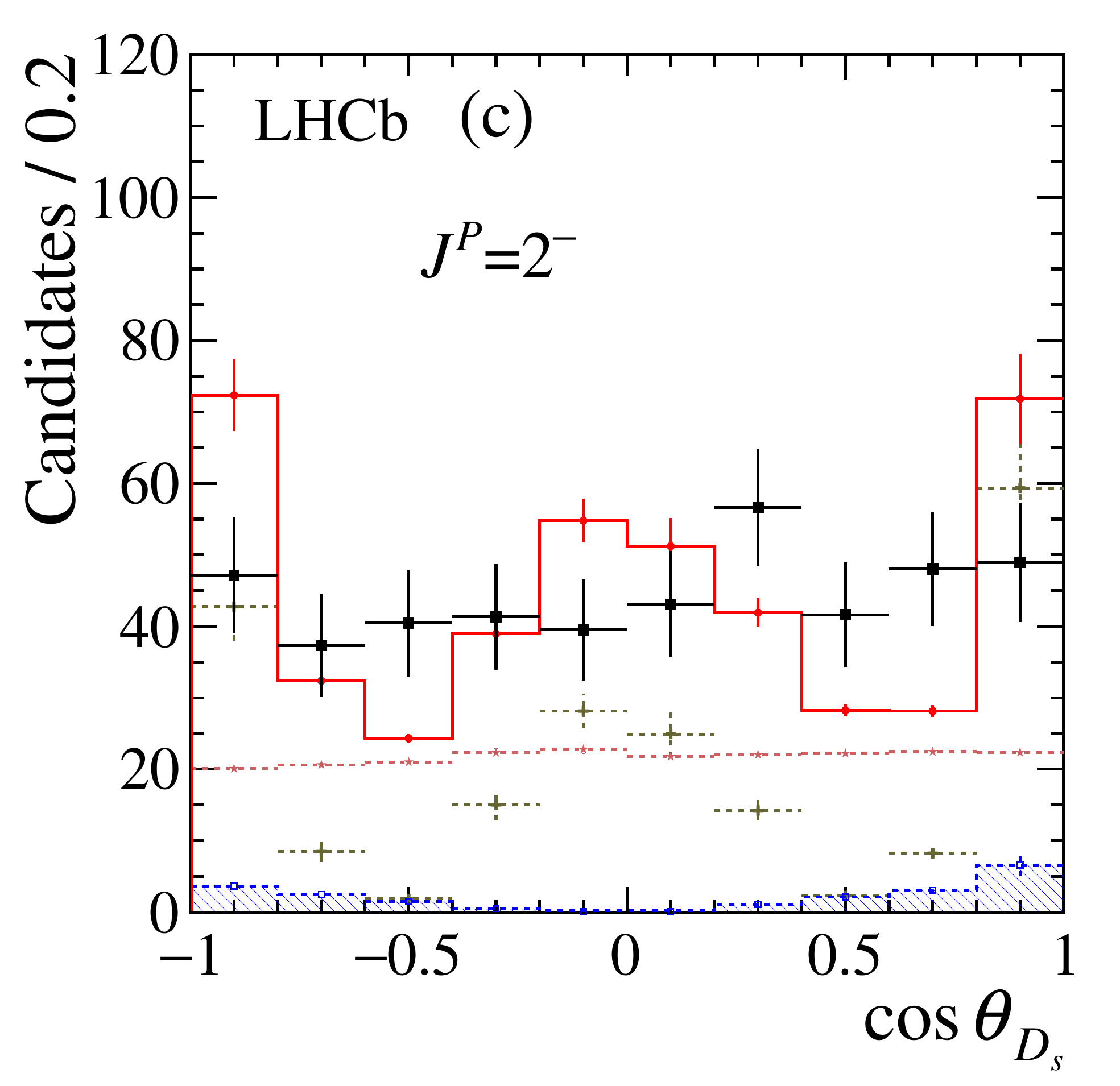}
    \caption{Comparison of $\cos \theta_{D_s}$ distributions for the spin parity of the \Dsn state assumed to be (a)~$0^-$, (b)~$1^+$, and (c)~$2^-$. 
    The $J^P=0^-$ model is the most consistent with data.  
    Data points are shown in black with the background subtracted statistically using the \sPlot method~\cite{Pivk:2004ty}. 
    Fit results are overlaid as a solid red histogram, with individual contributions shown as dotted histograms.}
    \label{fig:cos_Ds}
\end{figure}

Three possible spin parity models of the \DsJ state are tested: $J^P=0^-$, $1^+$ and $2^-$, among which the $J^P=0^-$ model leads to the best fit quality. 
This is understood by the property of $d^{J}_{0,0}(\theta_{D_s})$ describing the $\cos\theta_{D_s}$ behavior of the \DsJ state in the amplitude, which is proportional to the Legendre polynomial of order $J$ and is squared in the signal model.
Thus, the $\cos\theta_{D_s}$ distribution is described by a constant function for $J^P=0^-$, a second-order polynomial for $J^P=1^+$ and a fourth-order polynomial for $J^P=2^-$.
The $J^P=0^-$ model is clearly seen to be most consistent with data, as shown in Fig.~\ref{fig:cos_Ds}.

The significance to reject each of the disfavored spin parity models and the model without the \DsJ state is evaluated using the method of Refs.~\cite{LHCb-PAPER-2015-029,LHCb-PAPER-2016-019}. 
The difference of $-2\ln\Like(\Pars_{\rm min})$ between two models is assumed to follow a $\chi^2$ distribution.
The number of degrees of freedom associated with this $\chi^2$ distribution is one when comparing two models with different spin parity hypotheses, and twice the difference in the number of freely varying parameters when comparing models with different numbers of components. 
Taking the $J^P=0^-$ model as the reference, models without the \DsJ state or with $J^P=1^+$ or $2^-$ are all rejected with significance over ten standard deviations.
Therefore, the spin parity of the \DsJ state is determined to be $J^P=0^-$.
The \DsJ state is hereafter denoted as \Dsn. 

Almost equally good fit quality and the same \Dp\Kp\pim mass line shape are found for different width fractions $r$ in the range 0 to 1, indicating that this parameter cannot be determined using the current data.
The value of $r$ is fixed to $0.5$. 
The fitted BW mass and width of the \Dsn state vary significantly for different width fractions, but its pole position $\Mp-i\Gp/2$, where $\Mp$ and $\Gp$ are the pole mass and width, is found to be stable. 
This is understood as a consequence of the BW  parameters being dependent on specific reactions and width parameterizations, and as such having no strict physical meaning.
In contrast, the pole position is independent of the reaction studied and the chosen width parameterization, and is a physical characteristic of a resonance~\cite{PDG2020}. 
Therefore, only the pole mass and width of the \Dsn state are reported in this Letter.
These are measured to be $\Mp=2591\pm6\mev$ and $\Gp=89\pm16\mev$, where the uncertainty is statistical. 
Several mass projections are shown in Fig.~\ref{fig:projection}.
The enhancements in data at high $m_{\Dp\Dm}$ and low $m_{\Dp\pim}$ are seen to be well described as reflections of the \Dsn\ contribution.
A small excess is seen in $m_{\Dp\pim}$ near the mass of the $D_2^*(2460)^0$ state, which populates the region higher than $3\gev$ in \mDKpi, far away from the \Dsn peak. 
Therefore, vetoing it has small impact on the measured properties and is taken into account as a source of systematic uncertainty. Mass distributions of combinations of final-state particles not shown in Fig.~\ref{fig:projection} do not exhibit any structures.

Fit fractions, defined as in Ref.~\cite{LHCb-PAPER-2020-025}, for the three \Ds~components in the low \Kp\pim mass region obtained from the fit are listed in Table~\ref{tab:frac_sum}. The interference fraction between the \Dsn and NR components is also listed, whose negative central value explains why the full amplitude distribution lies below the NR distribution in some regions, as shown in Fig.~\ref{fig:projection}. The ratio of the \Dso and \Dsn fit fractions is also given in Table~\ref{tab:frac_sum}.

\begin{table}[btp]
    \centering
    \caption{Fit fractions for the three \Ds components in the low \Kp\pim mass region~(\mbox{$\mKpi<0.75\gev$}). 
     The interference fraction between the \Dsn and NR components is denoted as \Dszero-NR.
    There is no net interference between any other pair of components due to the orthogonality of the Wigner small-$d$ matrices for different spins.
    \Dsone/\Dszero denotes the ratio of the \Dso and \Dsn fit fractions.}
    \label{tab:frac_sum}
    \begin{tabular}{lr@{\:$\pm$\:}l@{\:$\pm$\:}l}\hline
         & \multicolumn{3}{c}{Fit fraction~($\times10^{-2}$)}  \\\hline
        \Dsn & $63\phdz$ & $\phz9\phdz\stat$ & $\phz9\phdz\syst$ \\
        \Dso & $3.9$ & $\phz1.4\stat$ & $\phz0.8\syst$ \\
        NR & $51\phdz$ & $11\phdz\stat$ & $19\phdz\syst$ \\
        \Dszero--NR  & $-18\phdz$ & $18\phdz\stat$ & $24\phdz\syst$ \\ \hline
        \Dsone/\Dszero & $6.1$ & $\phz2.4\stat$ & $\phz1.4\syst$ \\
        \hline
    \end{tabular}
\end{table}

\begin{table*}[btp]
    \centering
    \caption{Systematic uncertainties on the pole mass and width of the \Dsn state, and fit fractions of the three \Ds components. 
    The individual sources are added in quadrature to obtain the total uncertainty.
    The notations are the same as these in Table~\ref{tab:frac_sum}.}
    \label{tab:sys_sum}
    \begin{tabular}{l|cc|cccc|c}\hline
        Source & \multicolumn{1}{c}{$\Mp$} & \multicolumn{1}{c|}{$\Gp$} & \multicolumn{5}{c}{Fit fraction~($\times10^{-2}$)} \\ \cline{4-8} 
               & [${\rm MeV}$]  & [${\rm MeV}$] & $\Dszero$ & $\Dsone$ & NR & \Dszero--NR & \Dsone/\Dszero\\\hline
\Dsn width model & 6.1 & \phz8.0 & 4.7 & 0.0 & 15.0 & 19.6 & 0.5\\
\Dso mass shape & 0.3 & \phz4.3 & 2.3 & 0.6 & \phz3.5 & \phz5.3 & 1.1\\
\Kp\pim mass shape & 2.7 & \phz2.6 & 3.0 & 0.2 & \phz1.2 & \phz4.4 & 0.1\\
Blatt-Weisskopf factor & 0.7 & \phz3.4 & 2.8 & 0.3 & \phz1.3 & \phz3.0 & 0.2\\
Including \ccbar resonances & 1.1 & \phz5.4 & 2.7 & 0.1 & \phz6.3 & 10.0 & 0.4\\
\Dp\pim resonance veto & 2.4 & \phz2.1 & 4.6 & 0.3 & \phz9.4 & \phz4.6 & 0.2\\
Simulation correction & 0.2 & \phz1.1 & 0.3 & 0.1 & \phz0.7 & \phz0.8 & 0.2\\
Momentum calibration & 0.5 & \phz0.4 & 1.3 & 0.0 & \phz1.4 & \phz2.5 & 0.2\\
 \hline
Total &  7.2 &  11.7 &  8.6 &  0.8 &  19.3 &  23.9 &  1.4\\ \hline
    \end{tabular}
\end{table*}

Systematic uncertainties on the measured properties are summarized in Table~\ref{tab:sys_sum}. 
The primary source is related to the choice of the \Dsn width model, which is evaluated by describing the partial width of the ${\Dsn\to\D\kaon\pion}$ channel with a three-body formula similar  to that used in Ref.~\cite{LHCb-PAPER-2019-045}, instead of constant, or varying the width fraction $r$ between $0$ and $1$. 
Other sources include: variation of the \Dso mass shape due to uncertainties in the BW parameters and the width model, as well as the effect of detector resolution (which, at ${\cal O}(1\mev)$, is negligible for broader structures); the description of the \Kp\pim mass shape evaluated by using the LASS model~\cite{lass} and by varying within uncertainties the BW parameters of the \Kap state; variation of the Blatt-Weisskopf barrier factor, $R$, between $1.5$ and $4.5\gev^{-1}$; inclusion of possible \ccbar resonances, such as $\psi(3770)$, $\chi_{c0}(3930)$ and $\chi_{c2}(3930)$; vetoing possible $\Dp\pim$ resonant contributions by requiring $m_{\Dp\pim} < 2.4 \gev$; imperfections in the corrections applied to simulated events; and imperfect momentum calibration due to limited knowledge of the magnetic field and the detector alignment.
Uncertainties related to the size of the simulation sample are negligible.
The total systematic uncertainty is obtained by combining all contributions in quadrature.

In conclusion, a new excited \Ds meson is observed with large statistical significance in the \DKpi system of \BzDDKpi decays.
The analysis makes use of \proton\proton collision data collected by the LHCb experiment, corresponding to an integrated luminosity of 5.4\invfb. An amplitude analysis is performed on data in the low \Kpi mass region, $\mKpi<0.75\gev$, and the pole mass and width, and the spin parity of the new state are measured to be $\Mp=2591\pm6\pm7\mev$, $\Gp=89\pm16\pm12\mev$ and $J^P=0^-$, where the first uncertainty is statistical and the second systematic. Fit fractions obtained in the amplitude analysis are also reported. The new resonance, denoted as \Dsn, is a strong candidate to be the missing \DsS state, the radial excitation of the pseudoscalar ground-state \Ds\ meson.

\section*{Acknowledgements}
%
%
\noindent We express our gratitude to our colleagues in the CERN
accelerator departments for the excellent performance of the LHC. We
thank the technical and administrative staff at the LHCb
institutes.
We acknowledge support from CERN and from the national agencies:
CAPES, CNPq, FAPERJ and FINEP (Brazil); 
MOST and NSFC (China); 
CNRS/IN2P3 (France); 
BMBF, DFG and MPG (Germany); 
INFN (Italy); 
NWO (Netherlands); 
MNiSW and NCN (Poland); 
MEN/IFA (Romania); 
MSHE (Russia); 
MICINN (Spain); 
SNSF and SER (Switzerland); 
NASU (Ukraine); 
STFC (United Kingdom); 
DOE NP and NSF (USA).
We acknowledge the computing resources that are provided by CERN, IN2P3
(France), KIT and DESY (Germany), INFN (Italy), SURF (Netherlands),
PIC (Spain), GridPP (United Kingdom), RRCKI and Yandex
LLC (Russia), CSCS (Switzerland), IFIN-HH (Romania), CBPF (Brazil),
PL-GRID (Poland) and NERSC (USA).
We are indebted to the communities behind the multiple open-source
software packages on which we depend.
Individual groups or members have received support from
ARC and ARDC (Australia);
AvH Foundation (Germany);
EPLANET, Marie Sk\l{}odowska-Curie Actions and ERC (European Union);
A*MIDEX, ANR, Labex P2IO and OCEVU, and R\'{e}gion Auvergne-Rh\^{o}ne-Alpes (France);
Key Research Program of Frontier Sciences of CAS, CAS PIFI, CAS CCEPP, 
Fundamental Research Funds for the Central Universities, 
and Sci. \& Tech. Program of Guangzhou (China);
RFBR, RSF and Yandex LLC (Russia);
GVA, XuntaGal and GENCAT (Spain);
the Leverhulme Trust, the Royal Society
 and UKRI (United Kingdom).



\clearpage
\addcontentsline{toc}{section}{References}
\bibliographystyle{LHCb}
\bibliography{main,standard,LHCb-PAPER,LHCb-CONF,LHCb-DP,LHCb-TDR}

\newpage
\centerline
{\large\bf LHCb collaboration}
\begin
{flushleft}
\small
R.~Aaij$^{31}$,
C.~Abell{\'a}n~Beteta$^{49}$,
T.~Ackernley$^{59}$,
B.~Adeva$^{45}$,
M.~Adinolfi$^{53}$,
H.~Afsharnia$^{9}$,
C.A.~Aidala$^{84}$,
S.~Aiola$^{25}$,
Z.~Ajaltouni$^{9}$,
S.~Akar$^{64}$
% ...

{\footnotesize \it

% ...
$ ^{9}$Universit{\'e} Clermont Auvergne, CNRS/IN2P3, LPC, Clermont-Ferrand, France\\
% ...
$ ^{25}$INFN Sezione di Milano, Milano, Italy\\
% ...
$ ^{31}$Nikhef National Institute for Subatomic Physics, Amsterdam, Netherlands\\
% ...
$ ^{45}$Instituto Galego de F{\'\i}sica de Altas Enerx{\'\i}as (IGFAE), Universidade de Santiago de Compostela, Santiago de Compostela, Spain\\
% ...
$ ^{49}$Physik-Institut, Universit{\"a}t Z{\"u}rich, Z{\"u}rich, Switzerland\\
% ...
$ ^{53}$H.H. Wills Physics Laboratory, University of Bristol, Bristol, United Kingdom\\
% ...
$ ^{59}$Oliver Lodge Laboratory, University of Liverpool, Liverpool, United Kingdom\\
% ...
$ ^{64}$University of Cincinnati, Cincinnati, OH, United States\\
% ...
$ ^{67}$Syracuse University, Syracuse, NY, United States\\
% ...
$ ^{84}$University of Michigan, Ann Arbor, United States, associated to $^{67}$\\
\bigskip
% ...
\medskip
}
\end{flushleft}


\end{document}